\documentclass[fleqn,usenatbib]{mnras}
\usepackage[T1]{fontenc}
\usepackage{newtxtext,newtxmath}
\usepackage{graphicx}
\usepackage{amsmath}
\usepackage{booktabs}
\usepackage{url}

\title[Electron density from stacked JWST spectra]{Electron Densities of Typical Low-Mass Galaxies at $z\simeq$2--7 from Stacked JWST/NIRSpec Spectra}
\author[Liu \& Rong]{Shihong Liu$^{1,2}$, Yu Rong$^{1,2}$\thanks{Corresponding author; E-mail: rongyua@ustc.edu.cn}\\
$^{1}$Department of Astronomy, University of Science and Technology of China, Hefei 230026, China\\
$^{2}$School of Astronomy and Space Sciences, University of Science and Technology of China, Hefei 230026, Anhui, China\\
}

\begin{document}
\maketitle

\begin{abstract}

Direct electron-density measurements at high redshift are usually limited to galaxies with individually strong density-sensitive doublets, and therefore may not trace the average interstellar medium of ordinary low-mass galaxies. We stack public JWST/NIRSpec medium-resolution spectra from the DAWN JWST Archive to measure [SII]-based electron densities $n_e$ for low-mass galaxies at $2<z<7$. The stacks yield $n_e\simeq100$--$150\ {\rm cm^{-3}}$ at $2<z<5$ and $n_e=381^{+104}_{-89}\ {\rm cm^{-3}}$ at $5<z<7$, corresponding to an evolution $n_e=n_{e,0}[(1+z)/(1+2.3)]^\alpha$ with $n_{e,0}=76^{+22}_{-23}\ {\rm cm^{-3}}$ and $\alpha=1.88^{+0.60}_{-0.64}$. A mass-matched stacking test gives a consistent rising trend, indicating that the increase is not driven solely by changing stellar-mass distributions. Individually measurable galaxies with both [SII] components detected at ${\rm S/N}>5$ have a higher normalization, $n_{e,0}=211^{+36}_{-31}\ {\rm cm^{-3}}$, showing that individual-doublet samples select a denser subset. Stacking archival JWST spectra therefore provides a direct route to measuring the average gas density of low-mass galaxies below the individual-doublet detection threshold.

\end{abstract}
\section{Introduction}

The electron density of ionized gas, $n_e$, is a basic observable of HII regions and star-forming galaxies. It enters the emissivities of collisionally excited lines, affects the interpretation of strong-line metallicity and ionization diagnostics, and, together with the electron temperature, sets the thermal gas pressure $P/k\simeq n_eT_e$ \citep[e.g.][]{Osterbrock2006,Kewley2019}. The classical optical density diagnostics are forbidden-line doublets with different critical densities, especially [OII] $\lambda\lambda3726,3729$ and [SII] $\lambda\lambda6717,6731$ \citep{Osterbrock2006,Proxauf2014,Luridiana2015,Rong26}. In nearby star-forming galaxies these low-ionization diagnostics typically give densities of order $10$--$100\ {\rm cm^{-3}}$, whereas many high-redshift samples show values of several hundred ${\rm cm^{-3}}$ or higher \citep{Brinchmann2008,Liu2008,Sanders2016,Kaasinen2017,Kewley2019}.

Ground-based near-infrared spectroscopy established that the ionized gas in the redshift $z\simeq1$--3 galaxies is denser and more highly excited than in typical local galaxies, using [OII], [SII], and related rest-optical diagnostics \citep{Brinchmann2008,Liu2008,Shimakawa2015,Sanders2016,Masters2016,Kaasinen2017,Davies2021,Papovich2022}. These studies also showed that density is entangled with ionization parameter, gas-phase metallicity, star-formation-rate surface density, and galaxy structure \citep{Bian2016,Sanders2016,Kaasinen2017,Kewley2019,Reddy2023}. However, before JWST, the strongest rest-optical density diagnostics became difficult to observe beyond $z\simeq3$--4 from the ground, limiting both the redshift baseline and the ability to measure ordinary low-mass galaxies.

JWST/NIRSpec now provides medium-resolution rest-optical spectroscopy over $z>2$, enabling [OII] and [SII] density measurements deep into the epoch where galaxy sizes, gas fractions, star-formation intensities, and ionization conditions differ strongly from the local Universe \citep{Ferruit2022}. Early NIRSpec work used [OII] in $z\simeq4$--9 galaxies to infer elevated densities and an approximate $n_e\propto(1+z)^p$ evolution with $p\simeq1$--2 \citep{Isobe2023}. CEERS composite spectra at $z\simeq2.7$--6.3 showed that galaxies with larger O32 have higher [SII]-based densities and argued that gas density or star-formation-rate surface density may be central to regulating the ionization parameter \citep{Reddy2023}. The AURORA survey recently measured low-ionization densities from deep NIRSpec spectra, finding median values of $268_{-49}^{+45}$, $350_{-76}^{+140}$, and $480_{-310}^{+390}\ {\rm cm^{-3}}$ at $z=2.3$, 3.2, and 5.3 and a relation approximately following $(1+z)^{1.5\pm0.6}$ \citep{Topping2025}.

Most high-redshift density measurements are nevertheless conditioned on individual detections of weak density-sensitive doublets. This selection can favour galaxies with high equivalent widths, high excitation, high surface brightness, or compact star formation, and may not represent the mean conditions of the broader low-mass population \citep{Sanders2016,Reddy2023,Topping2025}. Stacking provides a complementary measurement: it sacrifices object-by-object information but can recover average line ratios for galaxies whose individual spectra do not yield reliable doublet densities. Public JWST archives make this approach particularly powerful because they combine many independent NIRSpec programmes over a wide redshift range.

In this paper we use public medium-resolution spectra from the DAWN JWST Archive\footnote{\url{https://dawn-cph.github.io/dja/}} \citep[DJA;][]{deGraaff24,Heintz23} to measure the average [SII]-based electron density of low-mass galaxies at $2<z<7$. We focus on [SII] because both lines are close in wavelength, are measured within a single grating for each object, and are better resolved than [OII] in our stacked spectra. We construct continuum-normalized stacks in redshift bins, fit the rest-frame 0.60--0.75 $\mu$m continuum window with stellar-population templates, measure [SII] $\lambda\lambda6717,6731$ with flexible line centres, compare our individual-spectrum measurements to the DJA emission-line table, and quantify the density--redshift trend.

\section{Data and Sample}

We use the DJA v4.4 \citep{deGraaff24,Heintz23}  public NIRSpec medium-resolution spectral products and emission-line catalog. The spectra were reduced with the public \texttt{msaexp} framework{\footnote{\url{https://zenodo.org/records/8319596}}} and distributed through the DAWN JWST Archive. The downloaded merged data products used here are
\texttt{dja\_msaexp\_emission\_lines\_v4.4.csv.gz} and the corresponding grating-filter spectral files. The primary [SII] analysis uses the medium-resolution G235M/F170LP and G395M/F290LP spectra. G235M covers [SII] for $2<z<4$, while G395M covers [SII] for $3<z<7$. We do not use PRISM spectra, and we do not use [OII] as a main diagnostic because the [OII] doublet is not reliably separated in the current stacks. This is consistent with the smaller wavelength separation of the [OII] doublet and the need for careful line-spread-function treatment in NIRSpec [OII] work \citep{Isobe2023}.

The stack parent sample is defined without using the strength of the [SII] doublet. We require a valid DJA spectral extraction, redshift $2<z<7$, redshift quality grade $\geq2$, finite stellar mass, $\log(M_\star/M_\odot)<9.5$, and conservative observed-frame coverage of the rest-frame 0.60--0.75 $\mu$m fitting interval, with a 0.01 $\mu$m margin from the nominal grating edges. These cuts define low-mass galaxies for which the [SII] region is observable, the stellar-continuum fit is well constrained on both sides of H$\alpha$ and [SII], and the rest-frame alignment is meaningful; they do not require [SII] to be detected in individual spectra. This is essential because the goal is to measure the average density of galaxies whose individual spectra are too weak for reliable [SII] doublet measurements.

The stacked bins are listed in Table~\ref{tab:stack}. The main analysis uses four accepted grating-redshift stacks: G235M/F170LP at $2<z<3$ and G395M/F290LP at $3<z<4$, $4<z<5$, and $5<z<7$. G235M/F170LP also covers [SII] for a small subset of the $3<z<4$ interval, but the fixed 0.60--0.75 $\mu$m rest-frame coverage requirement leaves only 123 spectra in that stack. Its stellar-continuum fit is visibly less stable and the resulting [SII] ratio lies beyond the low-density diagnostic limit, so we exclude it from the fiducial density-redshift analysis.

We also define an independent comparison sample of individually measurable galaxies. For this sample only, we require both fitted [SII] lines to have ${\rm S/N}>5$, a finite [SII]-based density $n_e$, and a fitted doublet separation exceeding the fitted full width at half maximum (FWHM). This sample is not used to define the stacked parent population. Instead, it provides a control set for verifying that our independently fitted [SII] line ratios and inferred electron densities agree with those derived from the DJA emission-line catalogue, and for showing where individually measurable systems lie relative to the stacks on the $n_e$--$z$ diagram. These two selections are deliberately different: the stacked sample measures the mean of the parent low-mass population, whereas the individual sample traces the subset whose density-sensitive doublet is already measurable in a single spectrum.

\begin{table*}
\centering
\small
\caption{Fiducial LSF-homogenized [SII] stack measurements. All spectra were smoothed to a common instrumental target of $\sigma_{\rm inst}=177.17\ {\rm km\,s^{-1}}$ before stacking, corresponding to an instrumental full width at half maximum ${\rm FWHM}_{\rm inst}=417.20\ {\rm km\,s^{-1}}$ and a resolving power $R\equiv\lambda/\Delta\lambda\simeq c/{\rm FWHM}_{\rm inst}\simeq719$ at [SII].}
\label{tab:stack}
\begin{tabular}{lcccccccc}
\toprule
Grating & $z$ bin & $N$ & $z_{\rm med}$ & $z_{16}$--$z_{84}$ & $\log M_\star$ & [SII] ratio & $n_e$ & S/N\\
 & & & & & & & ${\rm cm^{-3}}$ & 6717,6731\\
\midrule
G235M/F170LP & 2--3 & 659 & 2.52 & 2.17--2.88 & 8.78 & $1.351\pm0.021$ & $99^{+22}_{-20}$ & 107.8, 88.3\\
G395M/F290LP & 3--4 & 135 & 3.92 & 3.84--3.98 & 8.82 & $1.307\pm0.055$ & $151^{+66}_{-60}$ & 37.1, 32.6\\
G395M/F290LP & 4--5 & 603 & 4.48 & 4.14--4.82 & 8.66 & $1.329\pm0.033$ & $124^{+37}_{-33}$ & 68.6, 53.4\\
G395M/F290LP & 5--7 & 662 & 5.36 & 5.18--5.79 & 8.51 & $1.147\pm0.056$ & $381^{+104}_{-89}$ & 31.0, 29.2\\
\bottomrule
\end{tabular}
\end{table*}

\begin{figure*}
\centering
\includegraphics[width=0.92\textwidth]{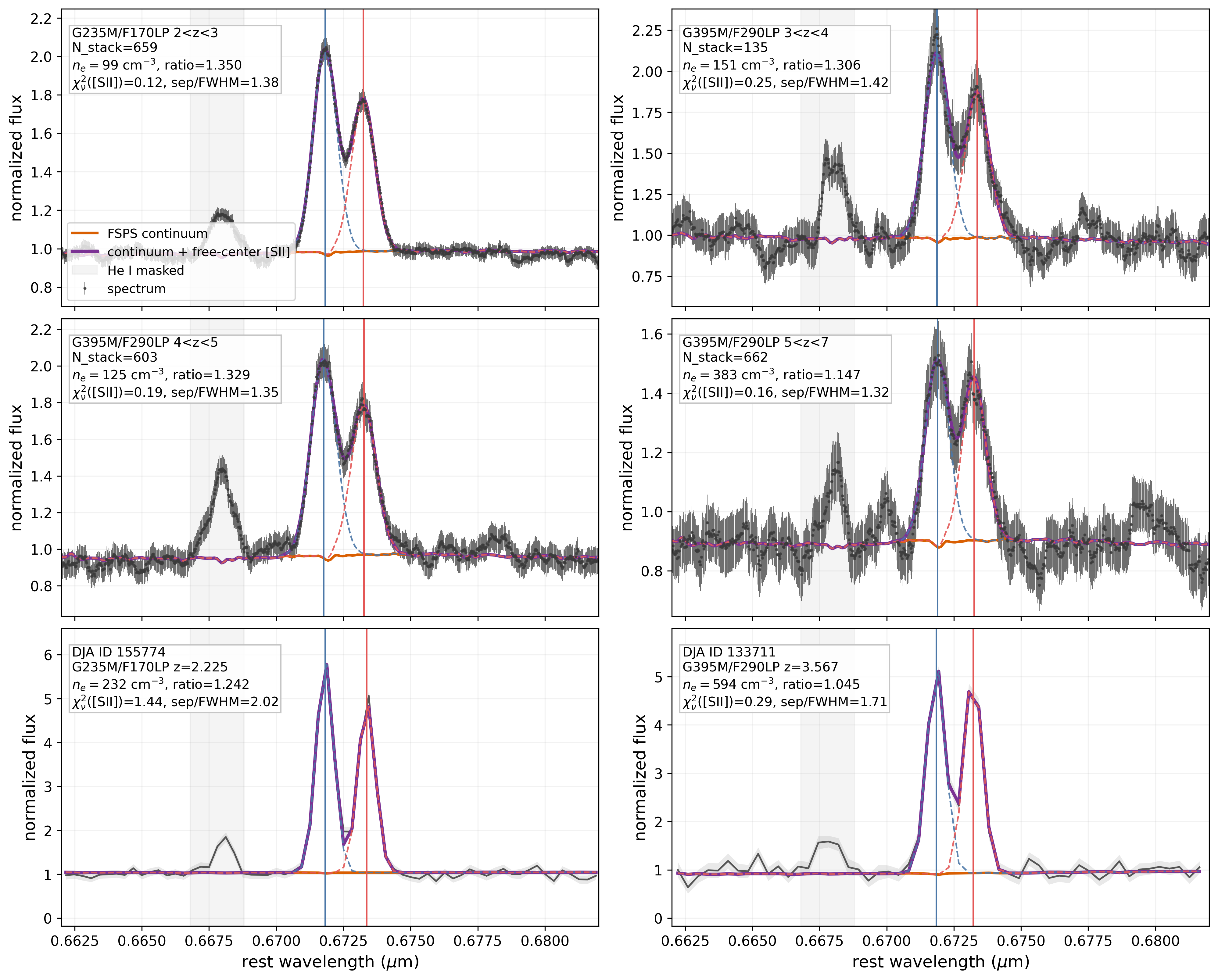}
\caption{Wide-window FSPS continuum fits and [SII] doublet fits. The four accepted stacked spectra are shown together with two representative high-S/N individual spectra. Black points or curves show the data, orange curves show the FSPS continuum model, and purple curves show the continuum plus the best-fitting [SII] double-Gaussian model. The blue and red dashed curves mark the continuum plus the [SII] $\lambda6717$ and $\lambda6731$ Gaussian components, respectively, and the vertical blue and red lines mark their fitted line centres. The plotted wavelength range is restricted to the [SII] region, but the continuum model is fit over the rest-frame 0.60--0.75 $\mu$m interval after masking emission lines.}
\label{fig:continuum}
\end{figure*}

\begin{figure}
\centering
\includegraphics[width=0.47\textwidth]{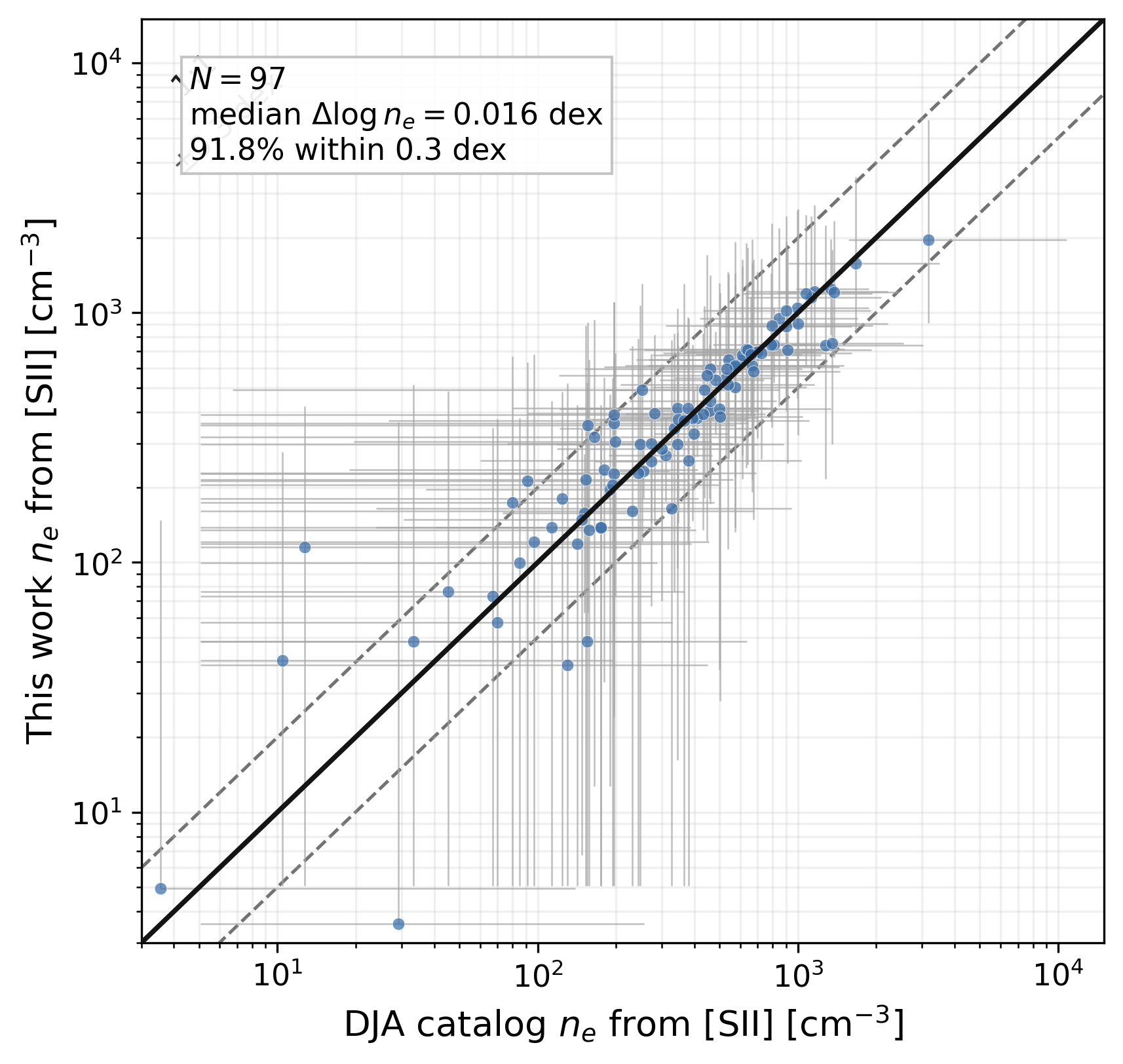}
\caption{Comparison between our individual-spectrum [SII] densities from wide-window FSPS continuum fitting and densities inferred from the DJA emission-line table. The comparison uses 97 objects with both fitted [SII] components detected at ${\rm S/N}>5$. The median offset is 0.016 dex.}
\label{fig:dja_compare}
\end{figure}

\section{Spectral Processing}

For each selected object we shift the observed spectrum to the rest frame using the DJA best redshift. We interpolate the spectrum onto a common rest-frame grid with spacing $2.5\times10^{-5}\ \mu$m. This spacing is a computational sampling grid, not the physical spectral resolution; the latter is set by the common LSF-homogenized instrumental FWHM described below. The adopted step samples the final [SII] instrumental FWHM by $\simeq37$ pixels and is used to avoid interpolation losses when median-combining spectra at different redshifts. Repeating the analysis with coarser grids does not change the inferred densities within the uncertainties, as shown in Section~\ref{discussion}. A key step is matching the spectral resolution before stacking. Because the NIRSpec line-spread function depends on grating and observed wavelength, the same rest-frame [SII] doublet is observed at different effective resolutions for galaxies at different redshifts. Directly stacking such spectra would mix different degrees of doublet blending and could bias the fitted [SII] line ratio. We therefore convolve each spectrum to a common instrumental resolution using the NIRSpec wavelength-dependent line-spread function appropriate for its grating and observed wavelength. Specifically, at the observed wavelength of [SII] for each object we estimate the instrumental Gaussian width $\sigma_{\rm inst}(\lambda_{\rm obs})$ and convolve the rest-frame spectrum by
\begin{equation}
\sigma_{\rm conv}=\left(\sigma_{\rm targ}^2-\sigma_{\rm inst}^2\right)^{1/2},
\end{equation}
where $\sigma_{\rm targ}=177.17\ {\rm km\,s^{-1}}$ is the maximum instrumental width among the stacks plus a 5 ${\rm km\,s^{-1}}$ safety margin. Thus the operation only smooths spectra to a common poorer resolution; no deconvolution is applied. The resulting common instrumental full width at half maximum is ${\rm FWHM}_{\rm inst}=2\sqrt{2\ln2}\,\sigma_{\rm targ}=417.20\ {\rm km\,s^{-1}}$. This corresponds to a resolving power $R\equiv\lambda/\Delta\lambda\simeq c/{\rm FWHM}_{\rm inst}\simeq719$ and to $\Delta\lambda_{\rm FWHM}=9.36$ \AA\ at the [SII] doublet.

For each object we build only the rest-frame 0.60--0.75 $\mu$m window used for this analysis. Before stacking, the spectrum is divided by a scalar continuum normalization measured from the two edge windows 0.600--0.625 $\mu$m and 0.725--0.750 $\mu$m. In each edge window we compute the median flux using valid pixels, and the final normalization is the median of the available edge-window medians. These edge windows are deliberately far from H$\alpha$, [NII], HeI, and [SII], so the normalization is not driven by the density-sensitive doublet. Similar continuum-scalar or UV-luminosity-scalar normalizations before constructing rest-frame composite spectra are commonly used to place high-redshift galaxy spectra on a comparable relative scale \citep[e.g.][]{Shapley2003,Topping2020,RobertsBorsani2024,Glazer2025,RobertsBorsani2026,Tang2026,Rong26a}. The individual continua are usually weak, as expected for low-mass high-redshift galaxies\footnote{In the four fiducial stacks the median DJA continuum S/N values are 0.46, 0.27, 0.24, and 0.21 from low to high redshift, and only 6--20\% of the contributing spectra have continuum S/N$>1$. After stacking, the median continuum S/N per stacked-spectrum pixel reaches about 8--30 in the two normalization edge windows.}. The broad edge windows are therefore used only to define a scalar relative normalization, not to fit individual stellar continua. We use median stacking with equal object weight as the fiducial method; the stack error spectrum is estimated from the Gaussian-equivalent object-to-object scatter divided by $\sqrt{N}$ at each wavelength.\footnote{We checked the source-to-source flux distributions at fixed wavelength in all four fiducial stacks. The distributions are approximately Gaussian in their central parts, the median and mean values are very close, and the central 16th--84th percentile intervals are nearly symmetric. We therefore use the Gaussian-equivalent standard deviation as the source scatter and divide this scatter by $\sqrt{N}$ (where $N$ is the number of sample) to define the stack error spectrum.} This choice intentionally avoids weighting the result toward the strongest emission-line objects, a known concern when comparing stacked spectra with individually detected emission-line samples \citep{Sanders2016,Reddy2023,Topping2025}.

The density measurement is performed after continuum subtraction on each stacked or individual spectrum. We fit the same rest-frame 0.60--0.75 $\mu$m window with FSPS stellar-population templates \citep{Conroy2009,Conroy2010} plus low-order additive continuum terms, while masking nebular emission lines including [OI], H$\alpha$, [NII], HeI, [SII], and [ArIII]. The residual [SII] profile is fit with two Gaussian components that share a common width. In Fig.~\ref{fig:continuum}, we show the fitting results for the four stacked spectra and two representative high-S/N individual spectra.

The fitted integrated flux ratio $F(6717)/F(6731)$ is converted to electron density using PyNeb \citep{Luridiana2015} at $T_e=10^4$ K. 
The [SII] temperature dependence is modest over the plausible high-redshift HII-region range \citep{Kewley2019,Topping2025}. The default values in Table~\ref{tab:stack} therefore quote the $T_e=10^4$ K conversion; the sensitivity to the assumed temperature is discussed in Section~\ref{discussion}.

For every stack or individual spectrum the central density is computed from the best-fitting line-flux ratio. The statistical uncertainty is then propagated from the fitted flux errors of the two [SII] components: we draw Gaussian realizations of $F(6717)$ and $F(6731)$ using their fitted $1\sigma$ errors, reject non-positive or physically invalid ratios outside the [SII] diagnostic range, convert the accepted ratios to $n_e$, and quote the 16th--84th percentile interval around the fixed best-fitting central value. The resulting $n_e$ uncertainties are therefore allowed to be asymmetric and are not symmetrized. In Fig.~\ref{fig:dja_compare}, we compare the densities inferred from our fitted integrated [SII] fluxes with those inferred from the DJA catalogue [SII] fluxes for the 97 high-S/N individual sources. The median difference is small, $\Delta \log n_e=0.016$~dex, indicating that our line fitting is on the same density scale as the independent DJA measurements.

\begin{figure*}
\centering
\includegraphics[width=0.92\textwidth]{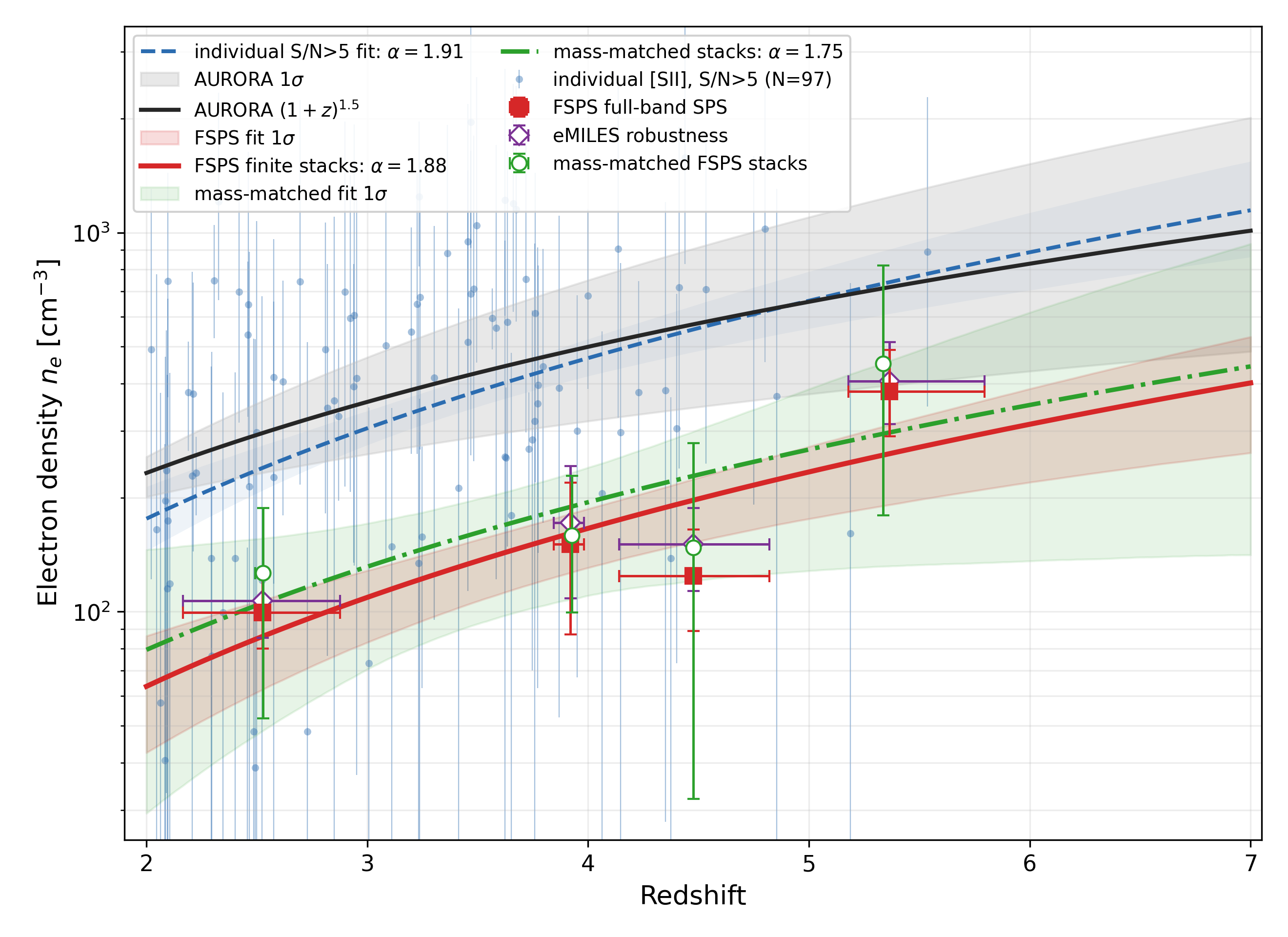}
\caption{Electron density versus redshift from [SII]. Blue circles are individually measurable galaxies with both fitted [SII] components detected at ${\rm S/N}>5$; the blue dashed line and pale band show their equal-weight bootstrap trend. Red squares are the accepted wide-window FSPS stack measurements of the parent low-mass sample; purple open diamonds show the eMILES-template robustness check. Green open circles show the mass-matched FSPS stacks. Red and green lines with shaded bands show the corresponding equal-weight stack fits and their 68\% Monte Carlo envelopes. Horizontal error bars on the stack points show the 16th--84th percentile redshift range of the contributing galaxies, not the full bin width. The black line and grey band show the AURORA reference relation, using $n_e(z=2.3)=268^{+45}_{-49}\ {\rm cm^{-3}}$ and slope $1.5\pm0.6$.}
\label{fig:evolution}
\end{figure*}

\section{Density Evolution}

Figure~\ref{fig:evolution} shows the individually measurable ${\rm S/N}>5$ comparison sample (light-blue dots) and the four accepted stacked measurements (red squares).
Following common parameterizations of density evolution in high-redshift emission-line studies \citep{Isobe2023,Topping2025}, we fit
\begin{equation}
n_e = n_{e,0}\left[\frac{1+z}{1+2.3}\right]^{\alpha},
\end{equation}
where $n_{e,0}$ and $\alpha$ are free parameters.
The normalization redshift $z=2.3$ is chosen to match the AURORA reference point \citep{Topping2025}. We perform an equal-weight least-squares fit to the measured point centres in $\log n_e$--$\log[(1+z)/(1+2.3)]$ space, to obtain the $\log n_e$--$z$ trend. To estimate the trend uncertainty, we perturb every $n_e$ within its asymmetric error distribution and the representative redshift within the 16th--84th percentile range of the galaxies contributing to that stack. The plotted 1$\sigma$ band is the 68\% Monte Carlo envelope around the central $\log n_e$--$z$ trend. Thus each galaxy or stack point contributes equally to the fitted population trend, while the measurement errors set the uncertainty.

The four accepted stacked points give
\begin{equation}
n_{e,0}=76^{+22}_{-23}\ {\rm cm^{-3}},\quad
\alpha=1.88^{+0.60}_{-0.64},
\end{equation}
as shown by the red line and shaded region in Fig.~\ref{fig:evolution}.
This slope is statistically consistent with the AURORA relation $\alpha=1.5\pm0.6$ \citep{Topping2025}. However, the absolute normalization is lower than AURORA (the black line and shaded region) and lower than the individually measurable sample (the blue dashed line and corresponding shaded region), because our stacks include many galaxies whose [SII] doublet is not individually measurable. This suggests that a large fraction of the low-mass high-redshift population has lower $n_e$ than the values inferred from samples selected by high-S/N [SII] detections.

To test whether the rising trend is driven by changing stellar-mass distributions across redshift, we repeat the full stacking and fitting procedure after matching the stellar-mass distributions. The reference distribution is the smallest accepted stack, G395M/F290LP at $3<z<4$. We split this reference sample into six stellar-mass quantile bins and, in every other redshift bin, draw the largest possible subset with the same relative occupancy across those mass bins. This keeps the reference stack intact and retains as many galaxies as possible in the other bins. The resulting stacks have median $\log(M_\star/M_\odot)=8.81$ in all redshift bins and use 492, 132, 420, and 378 spectra from low to high redshift. They give $n_e=127^{+61}_{-74}$, $159^{+69}_{-60}$, $148^{+130}_{-116}$, and $451^{+369}_{-271}\ {\rm cm^{-3}}$, respectively, as shown by the green open circles in Fig.~\ref{fig:evolution}. The fitted mass-matched $n_e$--$z$ relation has
\begin{equation}
n_{e,0}=94^{+58}_{-49}\ {\rm cm^{-3}},\quad
\alpha=1.75^{+1.25}_{-1.56}.
\end{equation}
The high-redshift stack remains the densest point after controlling the mass distribution. The rise in average density is therefore not explained solely by the lower median stellar mass of the highest-redshift parent sample.

The individual detections are not used to define the stacked trend. They occupy a broad density range and are shown as a comparison population because requiring both [SII] components at ${\rm S/N}>5$ selects the strongest-line systems. An equal-weight fit to the individual high-S/N detections gives $n_{e,0}=211^{+36}_{-31}\ {\rm cm^{-3}}$ and $\alpha=1.91^{+0.46}_{-0.45}$. This similar slope but higher normalization is consistent with the interpretation that individually measurable [SII] galaxies are a high-density subset rather than an unbiased average of the low-mass population.

\section{Discussion}
\label{discussion}

Our results support a picture in which the average low-ionization gas density of low-mass galaxies increases toward early cosmic times. The fiducial stacked slope, $\alpha=1.88^{+0.60}_{-0.64}$, is consistent with the $\alpha\simeq1$--2 evolution inferred from early JWST [OII] measurements \citep{Isobe2023} and with the AURORA low-ionization trend \citep{Topping2025}. The high-redshift stack at $z_{\rm med}=5.36$ is the strongest driver of the trend. The mass-matched test shows that this high-redshift enhancement persists when the stellar-mass distributions are forced to be nearly identical across the bins.

The lower normalization of the stacked relation compared to the individual detections is important. Individual [SII] density measurements in low-mass high-redshift galaxies preferentially select objects with strong, well-resolved doublets. These systems have typical densities of several hundred ${\rm cm^{-3}}$ and lie close to the normalization of other individually detected JWST density samples \citep{Reddy2023,Topping2025}. In contrast, the stacks include many spectra whose [SII] lines are too weak to measure individually and yield lower average densities at $z\simeq2$--5, rising sharply by $z\simeq5$--7. This suggests that the mean density of typical low-mass galaxies is not captured by individually detected doublet samples alone.
This offset can arise from both measurement selection and physical selection, and these two effects are difficult to separate in an individually detected comparison sample. First, the finite-density individual sample is selected to have strong, resolved [SII] doublets, so it naturally favours objects with high [SII] equivalent width and high line surface brightness \citep{Sanders2016,Reddy2023,Topping2025}. Second, the same selection may also be physical: strong [SII] emission and high surface brightness can preferentially select compact or intense star-forming regions, and high-redshift density measurements correlate with star-formation activity, ionization conditions, and star-formation-rate surface density \citep{Brinchmann2008,Sanders2016,Kaasinen2017,Reddy2023,Topping2025}. Therefore, the higher densities of the individually measurable galaxies should be interpreted as the density scale of a high-S/N, strong-line subset, not as the unbiased mean density of the parent low-mass population. The lower stacked densities are likely closer to the population average because the stacks include the many weak-[SII] systems that do not yield individual finite-density measurements.

We also tested whether the continuum normalization could be biased by the low per-pixel continuum S/N of the individual spectra. The normalization is not measured from a single pixel, but from the median continuum level across the two broad line-free windows used for the stack scaling. These windows contain about 2,000 pixels per object, so random pixel noise is partly averaged down in the scalar normalization. The median S/N values of the resulting normalization scalars are approximately 26, 19, 14, and 9 from the lowest to highest redshift bins, corresponding to median relative uncertainties of about 4\%, 5\%, 7\%, and 12\%. We further repeated the stack measurements without applying the continuum normalization. These no-normalization stacks still give low densities of order $100$--$200\ {\rm cm^{-3}}$ at $2<z<5$, and the $5<z<7$ stack reaches the low-density-limit regime. However, this test changes the meaning of the stack because galaxies with different luminosities/brightnesses are then combined directly, so the result can be affected by the luminosity/brightness distribution rather than representing an equal-object relative composite. We also considered photometric normalization, but bands near observed [SII] can be contaminated by strong H$\alpha$ emission, changing the effective weight of strong-line objects, while bands far from [SII] introduce colour and stellar-population dependence. We therefore do not consider photometric normalization or no normalization to be cleaner than the fiducial broad-window spectroscopic normalization for the present [SII] ratio measurement. Finally, removing the lowest 10--50\% of objects in continuum-normalization S/N and repeating the fiducial stack measurement does not move the stacks into the density regime of the high-S/N individual sample. We therefore keep the broad-window continuum-normalized median stack as the fiducial estimator, and these tests support the conclusion that the average parent population has lower electron densities than the strong-[SII] individual sample.

We performed an additional check of the individual-sample definition. Analogous to the work of \cite{Topping2025}, relaxing the 97-object high-S/N comparison sample from both [SII] components at ${\rm S/N}>5$ to at least one component at ${\rm S/N}>5$, while retaining finite density and resolved doublet requirements, increases the fitted individual sample to 128 objects. The relaxed sample gives $n_{e,0}=232^{+45}_{-37}\ {\rm cm^{-3}}$ and $\alpha=1.27^{+0.61}_{-0.62}$. A still looser both-component ${\rm S/N}>3$ selection gives 204 objects, with $n_{e,0}=268^{+37}_{-34}\ {\rm cm^{-3}}$ and $\alpha=0.99^{+0.42}_{-0.42}$. Both relaxed selections retain a higher normalization than the stacked relation, so the conclusion that individually measurable systems form a denser subset is not driven by the conservative both-line ${\rm S/N}>5$ definition.

We also tested whether the adopted rest-frame sampling grid biases the stacked line ratios. This test addresses the possibility that the fiducial $2.5\times10^{-5}\ \mu$m grid oversamples the native spectra and artificially sharpens or shifts the [SII] doublet. Repeating the complete wide-window stacking and FSPS fitting with coarser grids gives densities consistent with the fiducial values within the quoted uncertainties. For the four fiducial stacks, the $5\times10^{-5}\ \mu$m run gives $n_e=(111^{+31}_{-29},153^{+97}_{-80},113^{+51}_{-46},422^{+164}_{-132})\ {\rm cm^{-3}}$, while the $10^{-4}\ \mu$m run gives $n_e=(95^{+43}_{-40},154^{+140}_{-109},98^{+76}_{-66},407^{+233}_{-171})\ {\rm cm^{-3}}$. We therefore find no evidence that the finer fiducial sampling artificially drives the [SII] ratios or the inferred redshift trend.

We also performed a validation test to examine our continuum and line-fitting method. We fit the same [SII] region in the independent comparison sample of individual spectra, using the same wide-window FSPS continuum fitting and double-Gaussian model as for the stacks. We obtain 97 objects with both fitted [SII] components detected at ${\rm S/N}>5$ and finite densities in both our measurement and the DJA table. The median offset is $\Delta\log n_e=\log n_{e,\rm ours}-\log n_{e,\rm DJA}=0.016$ dex; 91.8\% of the common sample agrees within 0.3 dex. This demonstrates that the wide-window SPS-continuum method is not producing a large density-scale offset relative to the independent DJA emission-line catalogue.

We also examined whether the choice of stellar-population template affects the inferred [SII] densities. The fiducial analysis uses FSPS templates to model the stellar continuum across the rest-frame 0.60--0.75 $\mu$m window. As a robustness check, we repeated the same continuum-fitting, emission-line masking, and [SII] fitting procedure using eMILES templates \citep{Vazdekis2016}. The resulting densities, shown as purple open diamonds in Fig.~\ref{fig:evolution}, closely follow the fiducial FSPS stack measurements. For the four accepted stacks, eMILES gives $n_e=107$, 172, 151, and $406\ {\rm cm^{-3}}$, compared with the FSPS values of $99$, 151, 124, and $381\ {\rm cm^{-3}}$. These shifts are smaller than, or comparable to, the statistical uncertainties of the stacked points and do not change the inferred increase of average density toward higher redshift. We therefore treat the FSPS measurements as the fiducial values and use the eMILES points only as a continuum-template systematic check.

The density conversion also depends weakly on the assumed electron temperature. Table~\ref{tab:temperature} repeats the PyNeb conversion for the stacked [SII] ratios using $T_e=8000$--20,000 K, spanning a plausible range for high-redshift HII regions \citep{Kewley2019,Topping2025}. The low-density stacks change by less than or comparable to their statistical uncertainties. The highest-redshift stack is more temperature-sensitive because it lies at higher density, but it remains substantially denser than the $z\simeq2$--5 stacks across the full temperature range. The qualitative redshift trend is therefore not driven by the fiducial choice of $T_e=10^4$ K.

\begin{table}
\centering
\caption{Temperature dependence of stacked [SII] densities.}
\label{tab:temperature}
\begin{tabular}{lcccc}
\toprule
$z_{\rm med}$ & \multicolumn{4}{c}{$n_e\ [{\rm cm^{-3}}]$}\\
 & 8000 K & 10,000 K & 15,000 K & 20,000 K\\
\midrule
2.52 & 97 & 99 & 99 & 95\\
3.92 & 141 & 151 & 158 & 161\\
4.48 & 119 & 124 & 127 & 127\\
5.36 & 352 & 381 & 431 & 473\\
\bottomrule
\end{tabular}
\end{table}

Finally, our main stacked diagnostic uses only [SII]. This avoids cross-grating flux-calibration issues because both [SII] lines lie in the same grating for every object. In the stacked spectra, [OII] is detected but generally not resolved well enough for a robust stacked density ratio. High-S/N individual [OII] spectra are more nuanced: some approach useful resolution, but most require an explicit line-width or doublet-separation-to-FWHM quality cut before being included in a density analysis, as emphasized in recent JWST [OII] work \citep{Isobe2023}. Future work combining [OII], [SII], [ArIV], CIII], Balmer lines, and metallicity diagnostics will require explicit inter-grating flux calibration using overlapping continuum regions and strong emission lines \citep{Kewley2019,Topping2025}. For the present [SII] density ratio, that calibration is not required.

\section{Conclusions}

We have measured average [SII]-based electron densities for low-mass galaxies at $2<z<7$ by stacking public DJA JWST/NIRSpec medium-resolution spectra. Our main conclusions are:

\begin{enumerate}
\item The parent stacked sample is selected by redshift, stellar mass, spectral validity, and [SII] wavelength coverage, not by individual [SII] detectability; the accepted median stacks of 135--662 spectra per bin yield high-S/N [SII] $\lambda\lambda6717,6731$ detections with resolved doublet profiles.
\item In the individually measurable comparison sample, wide-window FSPS continuum fitting provides [SII] densities consistent with the DJA line table, with median offset 0.016 dex for 97 objects with both fitted [SII] components at ${\rm S/N}>5$.
\item The stacked low-mass sample follows $n_e=n_{e,0}[(1+z)/(1+2.3)]^\alpha$ with $n_{e,0}=76^{+22}_{-23}\ {\rm cm^{-3}}$ and $\alpha=1.88^{+0.60}_{-0.64}$ in an equal-weight fit to the accepted stack points.
\item A mass-matched stacking test gives a consistent trend,
$\alpha=1.75^{+1.25}_{-1.56}$, and leaves the $5<z<7$ stack as the highest-density point, indicating that the observed rise is not driven only by changing stellar-mass distributions.
\item The individually measurable galaxies have higher normalization, $n_{e,0}=211^{+36}_{-31}\ {\rm cm^{-3}}$, consistent with selection toward strong-line/high-density systems.
\end{enumerate}

The analysis demonstrates that stacking public JWST spectra can measure the average gas density of galaxies below the individual doublet-detection threshold. The most important next steps are field and observed-wavelength jackknife tests, refined grating systematics checks, and extension to pressure estimates using stacked temperature-sensitive diagnostics.

\section*{Acknowledgements}
The data products presented herein were retrieved from the Dawn JWST Archive (DJA). DJA is an initiative of the Cosmic Dawn Center (DAWN), which is funded by the Danish National Research Foundation under grant DNRF140. The analysis made use of NumPy, pandas, Astropy, SciPy, Matplotlib, and PyNeb. Y.R. acknowledges supports from the CAS Pioneer Hundred Talents Program (Category B), the NSFC grants 12522302 and 12273037, and the USTC Research Funds of the Double First-Class Initiative. This work is supported by the China Manned Space Program with grant no. CMS-CSST-2025-A06 and CMS-CSST-2025-A08.

\section*{Data Availability}
The JWST/NIRSpec spectra and emission-line catalogues used in this work are publicly available from the DAWN JWST Archive. The reduced stacks, measurement tables, and analysis scripts generated for this paper will be shared upon reasonable request and can be deposited in a public repository upon publication.

\end{document}